\documentclass{PoS}

\usepackage{amsmath}
\usepackage{amsthm}
\usepackage{amssymb}
\usepackage{amsfonts}

\newcommand{\Ort}{{\rm O\,}}

\newcommand{\USp}{{\rm USp\,}}
\newcommand{\U}{{\rm U\,}}
\newcommand{\SU}{{\rm SU\,}}

\newcommand{\eins}{\leavevmode\hbox{\small1\kern-3.8pt\normalsize1}}
\newcommand{\be}{\begin{eqnarray}}
\newcommand{\ee}{\end{eqnarray}}

\title{A classification of 2-dim Lattice Theory}

\ShortTitle{A classification of 2-dim Lattice Theory}

\author{\speaker{Mario Kieburg}\\
        Fakult\"at f\"ur Physik, Postfach 100131, 33501 Bielefeld, Germany\\
        E-mail: \email{mkieburg@physik.uni-bielefeld.de}}

\author{Jacobus J. M. Verbaarschot\\
        Department of Physics and Astronomy, State University of New York at Stony Brook, NY 11794-3800, USA\\
        E-mail: \email{jv@chi.physics.sunysb.edu}}

\author{Savvas Zafeiropoulos\\
        Department of Physics and Astronomy, State University of New York at Stony Brook, NY 11794-3800, USA and\\
        Laboratoire de Physique Corpusculaire, Universit\'e Blaise Pascal, CNRS/IN2P3 
63177 Aubi\`ere Cedex, France\\
        E-mail: \email{zafeiropoulos@clermont.in2p3.fr}}

\abstract{A unified classification and analysis is presented of two dimensional Dirac operators of QCD-like theories in the continuum as well as in a naive lattice discretization. Thereby we consider the quenched theory in the strong coupling limit. We do not only consider the case of a lattice which has  an even number of lattice sites in both directions and is thus equivalent to the  case of staggered fermions. We also study lattices with one or both directions with an odd parity to understand the general mechanism of changing the universality class via a discretization. Furthermore we identify  the corresponding random matrix ensembles sharing the  global symmetries of these QCD-like theories. Despite the Mermin-Wagner-Coleman theorem  we find good agreement of lattice data with our random matrix predictions.}

\FullConference{31st International Symposium on Lattice Field Theory LATTICE 2013\\
                 July 29 - August 3, 2013\\
                 Mainz, Germany}

\begin{document}
\section{Introduction}\label{intro}

Since the 90s random matrix theory (RMT) has been  successfully applied to four-dimensional QCD-like theories~\cite{V} and later also to three-dimensional ones~\cite{VZ3}. The reason for the good agreement of the spectral properties of both theories (the QCD-Dirac operator and chiral RMT)
in the microscopic limit is based on the fact that they share the same chiral Lagrangian. A chiral limit of QCD relies on the spontaneous breaking of chiral symmetry. In three and four dimensions this symmetry breaking is well understood. However in two dimensions the situation is less clear due to the Mermin-Wagner-Coleman theorem forbidding a spontaneous symmetry breaking of continuous symmetries in two and less dimensions. Nevertheless, numerical simulations \cite{num} as well as analytical examples \cite{anal} were given where the Mermin-Wagner-Coleman theorem does not seem to be applicable.

We discuss the classification of the continuum, see Sec.~\ref{sec2}, as well as the naive, see Sec.~\ref{sec3}, two dimensional Dirac operator in the microscopic domain. In particular we classify the Dirac operators via their global symmetries along the Cartan classification scheme \cite{class} and identify the corresponding RMTs. It is a well known fact that a naive discretization including staggered fermions generally exhibits a different universality class than the corresponding continuum limit \cite{Stag}. We confirm this for two-dimensional theories by  comparisons of lattice simulations to RMT predictions. The classification and analysis presented in this work is a summary of our work~\cite{KieVerZaf2013}.

\section{The classification of the 2-dim continuum Dirac operator}\label{sec2}

The symmetry classification of the 2-dim Euclidean Dirac operator,
\begin{eqnarray}\label{Diracdef}
 \mathcal{D}=\sigma_k\mathcal{D}_k=\left[\begin{array}{cc} 0 & \mathcal{W} \\ -\mathcal{W}^\dagger & 0 \end{array}\right]\quad {\rm with}\quad \mathcal{W}=\mathcal{D}_1+\imath\mathcal{D}_2\quad{\rm and}\quad \mathcal{D}_k=\partial_k+\imath A_k^a\lambda_a,
\end{eqnarray}
proceeds along the same lines as in four dimensions \cite{V} and in three dimensions \cite{VZ3}. The $2\times2$ matrices $\sigma_k$ are the Pauli matrices in spinor space and $\lambda_a$ are the generators for the gauge fields $A_\mu^a$ in a certain representation of the gauge group.

First, let us consider the gauge group $\SU(2)$ and the fermions in the fundamental representation. Thus the generators $\lambda_a$ are equal to the Pauli matrices $\tau_a$ acting in color space. Then the Dirac operator fulfils two symmetries,
\begin{equation}\label{symmetries-bet1}
 [\sigma_3,\mathcal{D}]_+=\sigma_3\mathcal{D}+\mathcal{D}\sigma_3=0\quad {\rm and} \quad \left([\tau_2\sigma_2K,\imath\mathcal{D}]_-=0\quad\Leftrightarrow\quad \mathcal{W}^T=-\tau_2\mathcal{W}\tau_2\right),
\end{equation}
where $K$ is the complex conjugation and $(.)^T\equiv K(.)^\dagger K$ is the transposition. Note that both symmetries are independently fulfilled. The first symmetry tells us that there exists a chiral basis. Since the anti-unitary operator fulfills $(\tau_2\sigma_2K)^2=1$ the second symmetry is equivalent with the fact that there is a basis for which the matrix elements of $\mathcal{D}$ become real. However, this basis is not a chiral basis since the matrix $\sigma_3$ does not commute with the anti-unitary operator, i.e. $[\tau_2\sigma_2K,\sigma_3]_-\neq 0$. This is  different from the four-dimensional case where $\gamma_5$ commutes with the anti-unitary symmetry of $\mathcal{D}$, see \cite{V}. Thus there is no way to find a chiral basis such that the two-dimensional Dirac operator becomes real. Instead we find a condition for $\mathcal{W}$, see Eq.~\eqref{symmetries-bet1}, telling us that $\mathcal{W}$ is complex anti-selfdual or equivalently $\tau_2\mathcal{W}$ is complex symmetric. Ignoring at the moment the Mermin-Wagner-Coleman theorem, the RMT corresponding to  the Dirac operator is obtained by replacing
the  operator $\tau_2\mathcal{W}$ by a complex symmetric, Gaussian distributed random matrix $W$. 
According to universality arguments, the Dirac operator $\mathcal{D}$ would not only share the symmetry breaking pattern, $\USp(2N_{\rm f})\times\USp(2N_{\rm f})\rightarrow\USp(2N_{\rm f})$\footnote{$N_{\rm f}$ is the number of flavors.}, but also the spectral properties of this RMT in particular the linear level repulsion (Dyson index $\beta=1$) and the linear repulsion from the origin (for the quenched case). The corresponding RMT is one of the two Boguliubov-deGennes ensembles denoted by CI in the Cartan classification \cite{class}, see table~\ref{table1}.

\begin{table}
\begin{tabular}{|c|c|c|c|}\hline
\quad Dimension\quad & \quad Dyson index $\beta $\quad & Symmetry Breaking Pattern & RMT \\
\hline\hline
2   & 1  &    \quad$\USp(2N_{\rm f}) \times \USp(2N_{\rm f}) \to \USp(2N_{\rm f})$\quad &  (CI) \\  
2   & 2  &    $\U(N_{\rm f})\times \U(N_{\rm f}) \to \U(N_{\rm f})$& \quad chGUE (AIII)\quad \\
2   & 4  &    $\Ort(2N_{\rm f}) \times \Ort(2N_{\rm f}) \to \Ort(2N_{\rm f})$&   (DIII) \\
\hline
3  & 1   &$\USp(4N_{\rm f}) \to \USp(2N_{\rm f}) \times \USp(2N_{\rm f})$ & GOE   (AI)\\ 
3  & 2   &$\U(2N_{\rm f}) \to \U(N_{\rm f}) \times \U(N_{\rm f})$   &GUE (A)\\
3  & 4   & $\Ort(2N_{\rm f}) \to \Ort(N_{\rm f}) \times \Ort(N_{\rm f})$& GSE (AII)\\
\hline
4  & 1 &   $\U(2N_{\rm f})/\USp(2N_{\rm f})$ & chGOE (BDI)\\
4  & 2 &   $\U(N_{\rm f})\times \U(N_{\rm f}) \to \U(N_{\rm f})$ & chGUE (AIII)\\
4  &  4 &  $\U(2N_{\rm f})/ \Ort(2N_{\rm f})$ & chGSE  (CII)\\ \hline
\end{tabular}
 \caption[]{The chiral symmetry breaking pattern in two, three, and four dimensions  for 
different values of the Dyson index $\beta$ with the corresponding RMT classified via the Cartan scheme \cite{class}. It agrees with the more general classification for an arbitrary dimension presented in \cite{bott} where the Bott periodicity of this classification has been shown.
\label{table1}}
\end{table}

For the gauge groups $\SU(N_{\rm c})$ with more than two colors, $N_{\rm c}>2$, and the fermions in the fundamental representation only the chiral symmetry survives for the Dirac operator $\mathcal{D}$,
\begin{equation}\label{symmetries-bet2}
 [\sigma_3,\mathcal{D}]_+=0.
\end{equation} Thus the corresponding random matrix is the one of the four-dimensional theory for these gauge theories, namely $\mathcal{W}$ is replaced by a complex, Gaussian distributed random matrix $W$. The level repulsion is quadratic ($\beta=2$)  and the repulsion of the eigenvalues of $\mathcal{D}$ from the origin is at least linear. The RMT is the well-known chiral Gaussian Unitary Ensemble (chGUE) denoted by the Cartan symbol AIII and with the symmetry breaking pattern $\U(N_{\rm f})\times\U(N_{\rm f})\rightarrow\U(N_{\rm f})$, see table~\ref{table1}.

The third class of $\SU(N_{\rm c})$ gauge theories we consider are those with the fermions in the adjoint representation. Then the generators $\lambda_a$ are purely imaginary anti-symmetric matrices. As in the case $\beta=1$ we find two symmetries fulfilled by $\mathcal{D}$,
\begin{equation}\label{symmetries-bet4}
 [\sigma_3,\mathcal{D}]_+=0\quad {\rm and} \quad \left([\sigma_2K,\imath\mathcal{D}]_-=0\quad\Leftrightarrow\quad \mathcal{W}^T=-\mathcal{W}\right).
\end{equation}
The first one is again equivalent to the existence of a chiral basis while the second one refers to the existence of a basis for which the Dirac operator becomes quaternion real as in the four-dimensional theory \cite{V} because the anti-unitary symmetry fulfils $(\sigma_2K)^2=-1$. However, in contrast to the four-dimensional theory,  the chiral symmetry operator does not commute with the anti-unitary one. Hence the second symmetry yields a condition on $\mathcal{W}$ telling us that $\mathcal{W}$ can be replaced by a complex anti-symmetric, Gaussian distributed random matrix $W$.  Its level repulsion is quartic  ($\beta=4$) and exhibits a linear or a quintic repulsion from the origin depending on whether $W$ is even or odd dimensional, respectively. Moreover the eigenvalues are all Kramers degenerate. The RMT is the other Bogoliubov-deGennes ensemble denoted by DIII \cite{class} and shares the same symmetry breaking pattern, $\Ort(2N_{\rm f})\times\Ort(2N_{\rm f})\rightarrow\Ort(2N_{\rm f})$, with the Dirac operator $\mathcal{D}$, see table~\ref{table1}.

\section{The classification of the 2-dim naive Dirac operator}\label{sec3}

Introducing a two-dimensional, periodic $L_1\times L_2$ lattice the situation drastically changes since new symmetries may arise depending on whether the number of lattice sites in some directions, $L_{1/2}$, is odd or even. The covariant derivatives $\mathcal{D}_k$, see Eq.~\eqref{Diracdef}, are replaced by the naive discretization scheme. Furthermore the symmetries~(\ref{symmetries-bet1}-\ref{symmetries-bet4}) still hold for the corresponding gauge theories but do not exclude additional symmetries which may result from the discretization. Moreover the ensuing discussion does not incorporate the  statistical weight from the Wilson action. Hence we consider only the limit of strong coupling where the gauge group elements are generated by the corresponding Haar measure of the gauge group.

\begin{table}
\begin{tabular}{|c|c|c|c|c|c|c|}\hline
$\beta$ & Lattice sites & Sym. Class & $\beta_{\rm eff}$&  $\alpha$ & Degen. & Symmetry Breaking Pattern \\
\hline\hline
1   &   ee  & CII & 4&3 & 4& $\U(2N_{\rm f}) \to \Ort(2N_{\rm f})$\\
1   &   eo  & C  &2 & 2 & 2 &$\USp(4N_{\rm f}) \to \U(2N_{\rm f})$     \\
1   & oo  & CI & 1&1 &1 &  $\USp(2N_{\rm f}) \times \USp(2N_{\rm f}) \to \USp(2N_{\rm f})$\\
\hline
2   & ee  &AIII  & 2& 1 & 2& $\U(2N_{\rm f})\times \U(2N_{\rm f}) \to \U(2N_{\rm f})$\\
2   & eo  & A & 2& 0 & 2 & $\U(2N_{\rm f})\to \U(N_{\rm f}) \times \U(N_{\rm f})$ \\
2   & oo &AIII  & 2 & 1 & 1&  $\U(N_{\rm f})\times \U(N_{\rm f}) \to \U(N_{\rm f})$ \\
\hline
4   & ee  & BDI & 1& 0 & 2& $\U(4N_{\rm f}) \to \USp(4N_{\rm f}) $ \\
4   & eo  & D  & 2&  0& 2 & $\Ort(4N_{\rm f}) \to \U(2N_{\rm f})$\\
4   & oo  & DIII (even-dim)  & 4 & 1 &2 &  $\Ort(2N_{\rm f}) \times \Ort(2N_{\rm f}) \to \Ort(2N_{\rm f})$\\
4   & oo  & DIII (odd-dim)   & 4 & 5 & 2& $\Ort(2N_{\rm f}) \times \Ort(2N_{\rm f}) \to \Ort(2N_{\rm f})$ \\ \hline
\end{tabular}
\caption[]{Corresponding RMTs for all two dimensional naive fermions of QCD-like theories. The Dyson index $\beta$ 
refers to the anti-unitary symmetry of the continuum Dirac operator while $\beta_{\rm eff}$ indicates the level repulsion for the naive Dirac operator. The parameter $\alpha$ is the generic repulsion of the eigenvalues from the origin. Depending on if we have an even or odd number of lattice sites, the symmetry classes according to the ten-fold classification of RMTs \cite{class} as well as the generic degeneracy (sixth column) of the eigenvalues change drastically. In the last column the symmetry breaking patterns are shown, respectively.
\label{table2}}
\end{table}

Assuming an even parity in the $x$-direction ($L_1$ even) we may define an operator $\Gamma_5^{(1)}$ assigning a ``$+$''-sign to an even lattice site and a  ``$-$''-sign to an odd one. This operator fulfils the relations
\begin{equation}\label{commrel}
\Gamma_5^{(1)}\mathcal{D}_1\Gamma_5^{(1)}=-\mathcal{D}_1,\ \Gamma_5^{(1)}\mathcal{D}_2\Gamma_5^{(1)}=\mathcal{D}_2\quad \Rightarrow\quad [\Gamma_5^{(1)}\sigma_2,\mathcal{D}]_-=0.
\end{equation}
We define the unitary rotation $\Pi_5^{(1)} \equiv \exp[\imath\pi(\eins  -\Gamma_5^{(1)})/4]$. Then Eq.~\eqref{commrel} is equivalent to
\begin{equation}\label{commrela}
(\Pi_5^{(1)} \mathcal{W}\Pi_5^{(1)} )^\dagger=\Pi_5^{(1)} \mathcal{W}\Pi_5^{(1)} ,
\end{equation}
i.e. $\Pi_5^{(1)} \mathcal{W}\Pi_5^{(1)} $ is Hermitian. A similar relation can be found in the case if the lattice has an even number of lattice sites in the $y$-direction, namely
\begin{equation}\label{commrelb}
[\Gamma_5^{(2)}\sigma_1,\mathcal{D}]_-=0\quad \Rightarrow\quad (\imath\Pi_5^{(2)} \mathcal{W}\Pi_5^{(2)} )^\dagger=\imath\Pi_5^{(2)} \mathcal{W}\Pi_5^{(2)} ,
\end{equation}
implying that $\imath \Pi_5^{(2)} {\mathcal W}   \Pi_5^{(2)}$ is Hermitian. These new symmetries have to be combined with the one of the continuum Dirac operator, see Eqs.~(\ref{symmetries-bet1}-\ref{symmetries-bet4}), yielding new symmetry classes.

First, let us consider an odd number of lattice sites in both directions. Then  nothing changes from the symmetry discussion in Sec.~\ref{sec2}. Also the corresponding RMTs will be the same, cf. table~\ref{table2}. This is confirmed by lattice simulations performed at a small volume ($L_{1/2}<10$), see Figs.~1a, 1e and 1f for $\beta = 1, 4$ and 2, respectively. The case $\beta=4$ is  peculiar. Since $\mathcal{W}$ is antisymmetric it may have an additional generic zero eigenvalue if this operator is odd-dimensional in contrast to an even dimensional $\mathcal{W}$. Thus depending on the number of colors the repulsion of the eigenvalues of $\mathcal{D}$ from the origin changes drastically.

\begin{figure}[!ht]
\includegraphics[width=0.5\textwidth]{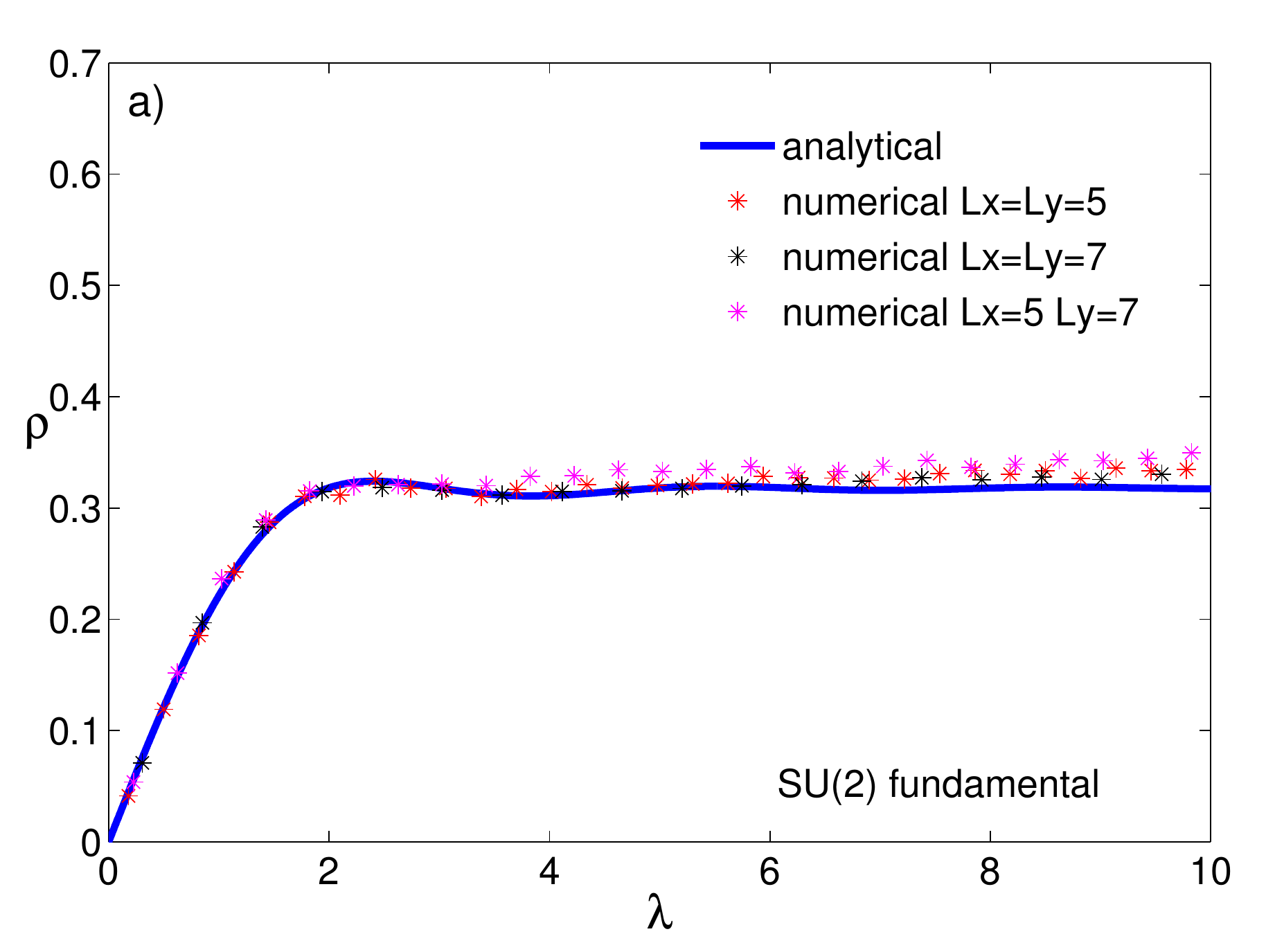}\includegraphics[width=0.5\textwidth]{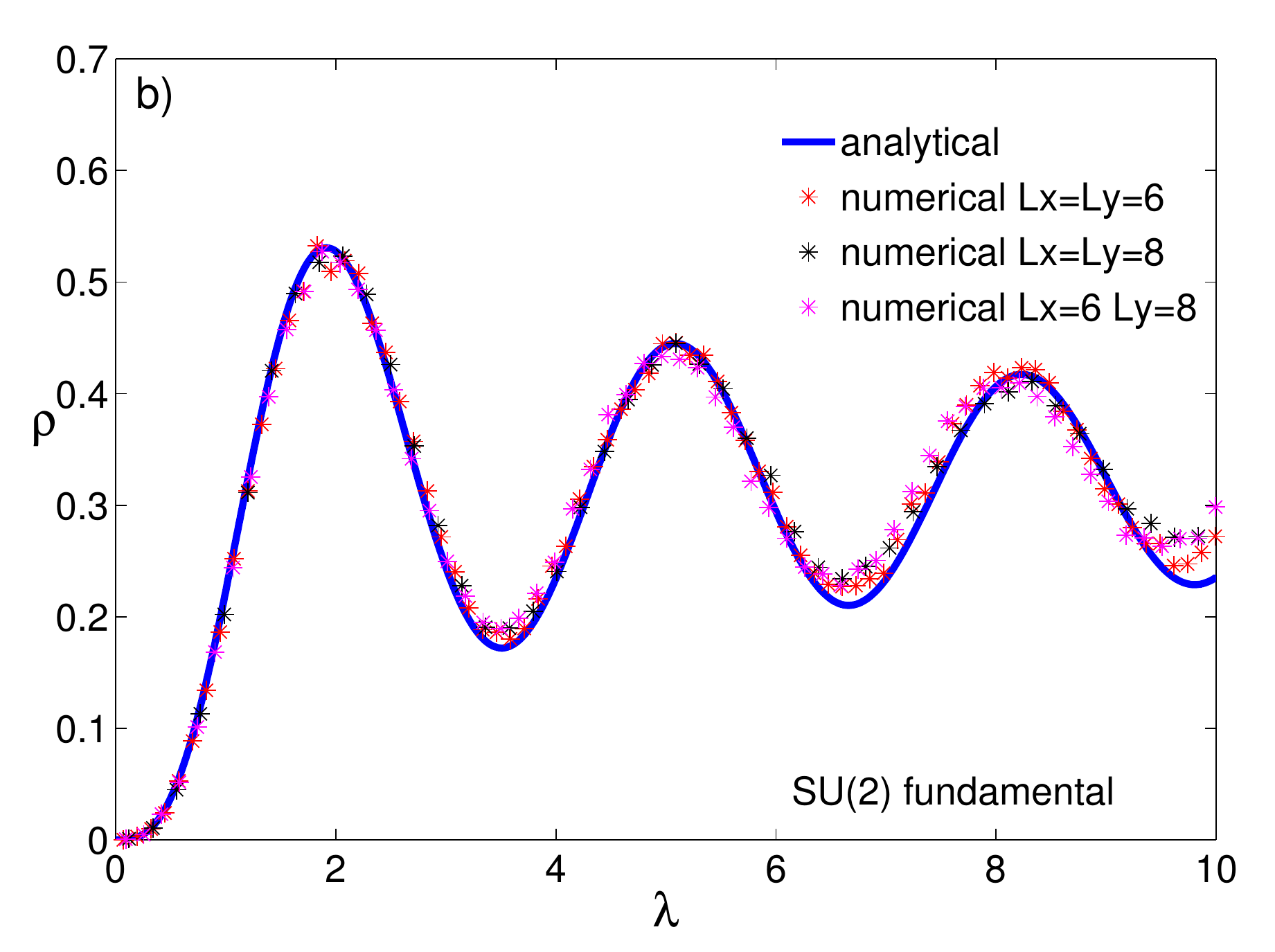}\\
\includegraphics[width=0.5\textwidth]{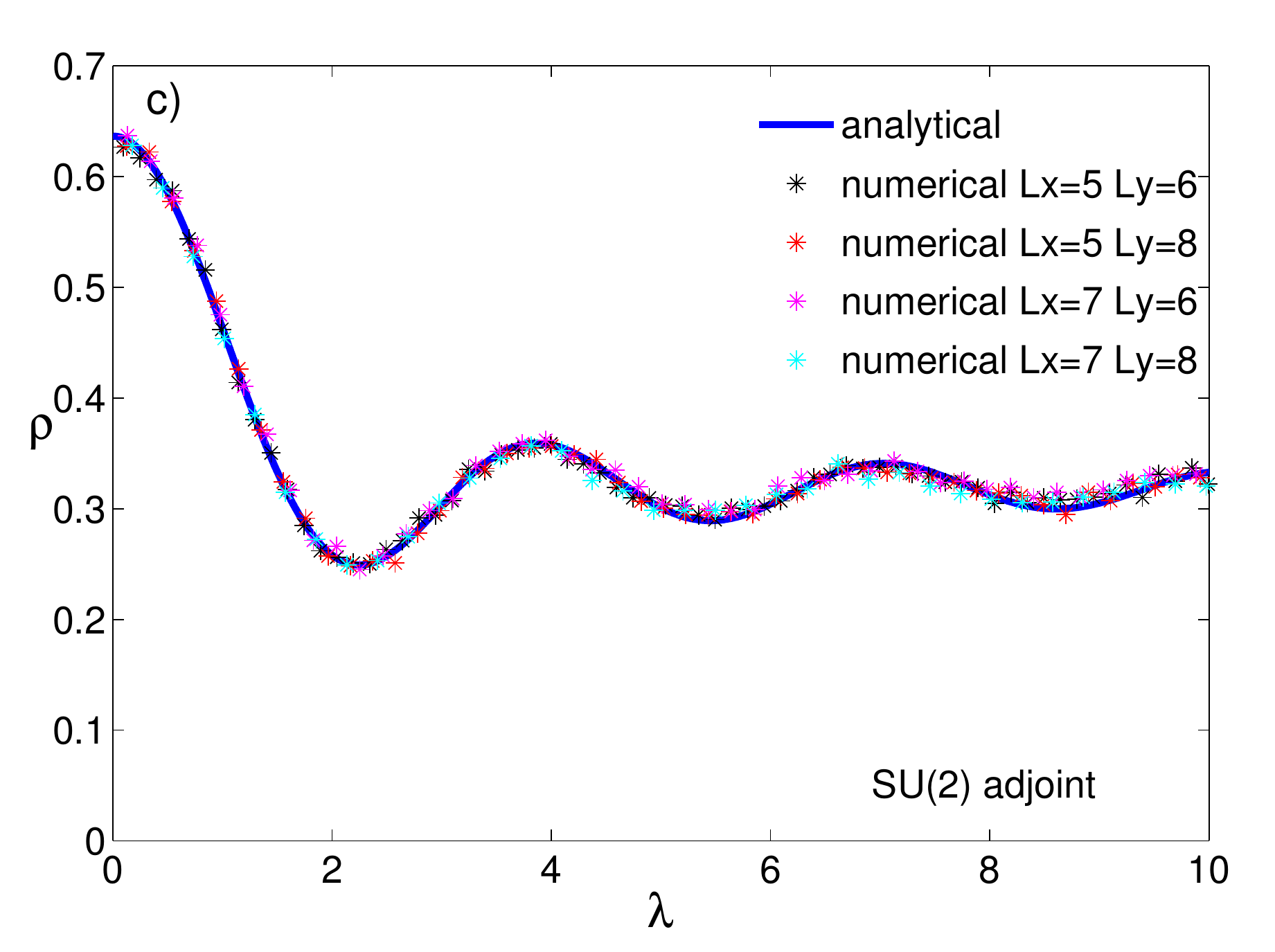}\includegraphics[width=0.5\textwidth]{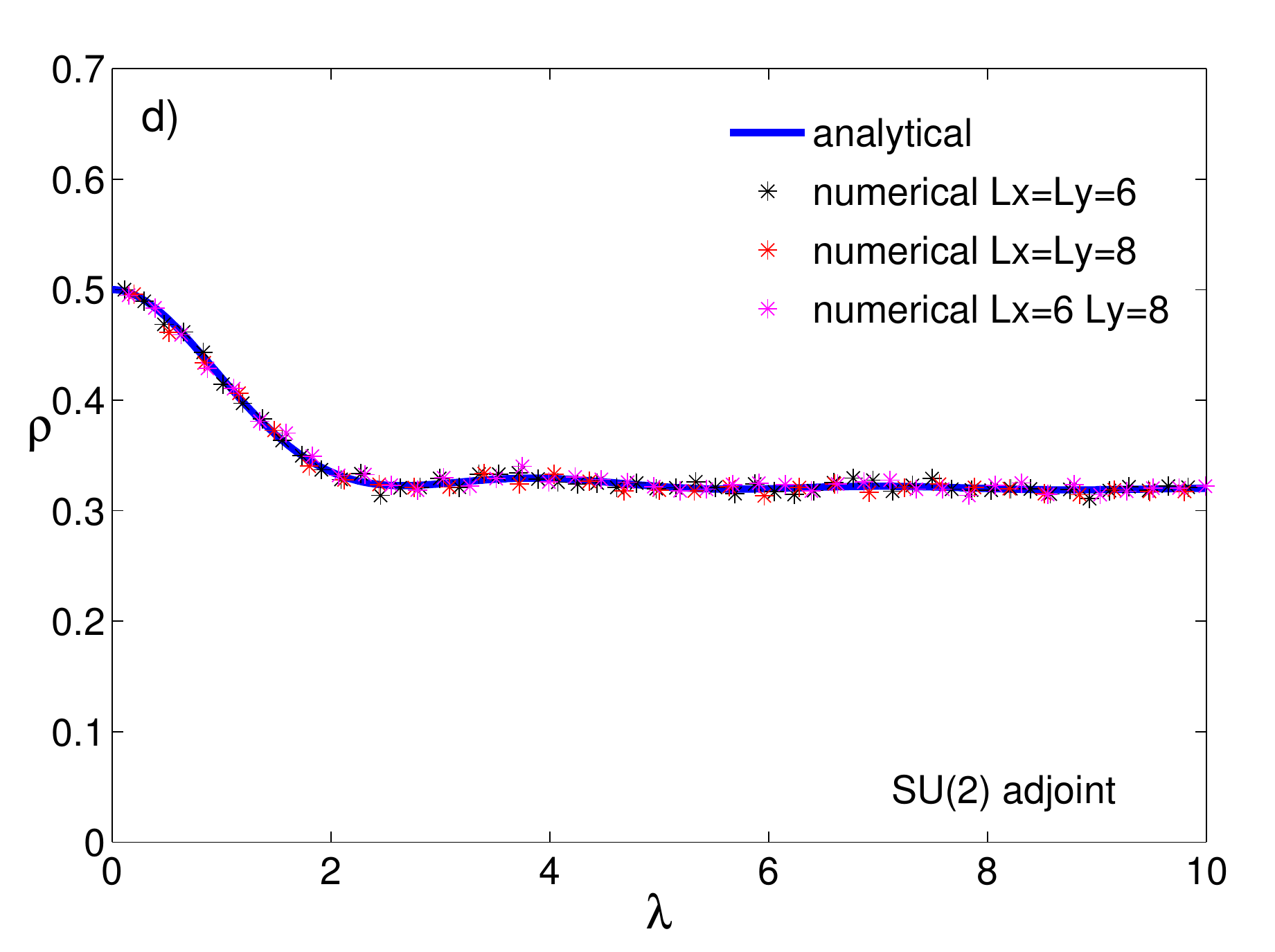}\\
\includegraphics[width=0.5\textwidth]{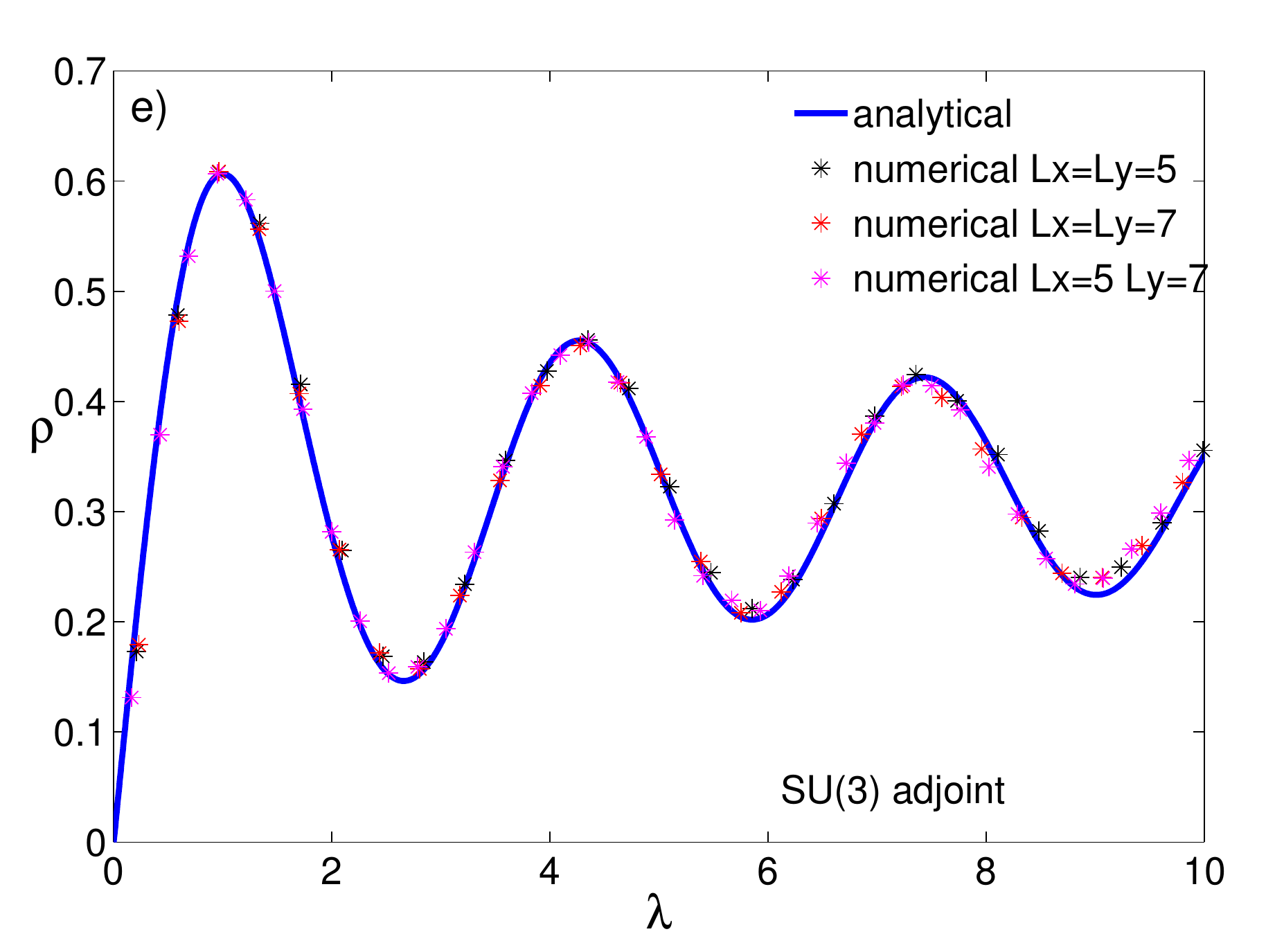}\includegraphics[width=0.5\textwidth]{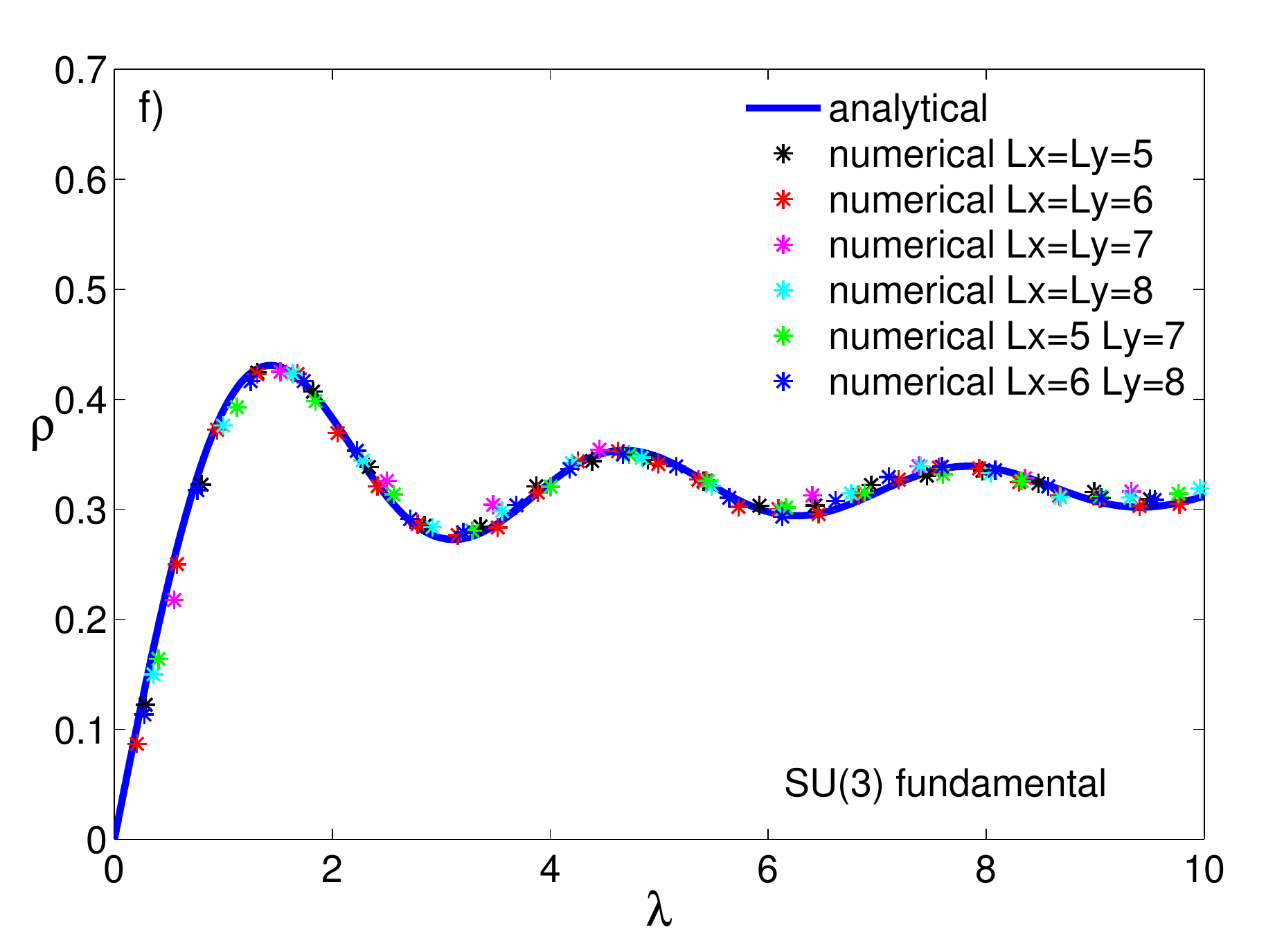}
\caption{Comparisons of some RMT predictions (curves) with quenched lattice simulations of naive fermions (crosses). Notice that the lattice data were generated by Monte Carlo simulations in the strong coupling limit, meaning that the gauge group elements on the links were drawn from the Haar-measure of the gauge group, only. Shown are the microscopic level densities as follows: fundamental $\SU(2)$ on an odd-odd lattice (a), fundamental $\SU(2)$ on an even-even lattice (b),  adjoint $\SU(2)$ on an even-odd lattice (c),  adjoint $\SU(2)$ on an even-even lattice (d),   adjoint $\SU(3)$ on an odd-odd lattice (e) and fundamental $\SU(3)$ on an even-even and odd-odd lattice (f). Notice that the even-even case  corresponds to the staggered Dirac operator. 
 The microscopic level densities are normalized such that it asymptotes to the  value of $1/\pi$.}
\label{fig1}
\end{figure}

In the case of a mixed situation ($L_1+L_2$ odd) the symmetries completely change. Assuming $L_1$ to be odd, the new symmetry combined with Eqs.~(\ref{symmetries-bet1}-\ref{symmetries-bet4}) tells us that $\Pi_5^{(1)} \mathcal{W}\Pi_5^{(1)} $ is first of all Hermitian for all three Dyson indices $\beta=1,2,4$. Furthermore, $\Pi_5^{(1)} \mathcal{W}\Pi_5^{(1)} $ is anti-selfdual for $\beta=1$, constructable by random matrices in the Lie-algebra ${\rm sp}(2N)$, and is anti-symmetric for $\beta=4$ which corresponds to random matrices in the Lie-algebra ${\rm o}(N)$. The splitting of the universality class in an even and odd dimensional $\mathcal{W}$ for $\beta=4$ does not appear in this case since the matrix dimension of $\mathcal{W}$ is proportional to $L_1L_2$ and, hence, is always even. A full classification is listed in table~\ref{table2} and some comparisons with lattice data are shown in Fig.~\ref{fig1}c.

In the last case ($L_1$ and $L_2$ are even) both symmetry relations~\eqref{commrela} and \eqref{commrelb} apply. Equation~\eqref{commrela} combined   with Eqs.~(\ref{symmetries-bet1}-\ref{symmetries-bet4}) ensures that $\Pi_5^{(1)} \mathcal{W}\Pi_5^{(1)} $ is still Hermitian for all three $\beta$ and anti-selfdual for $\beta=1$ and anti-symmetric for $\beta=4$. Additionally, combining Eqs.~\eqref{commrela} and \eqref{commrelb}, the operator $\Pi_5^{(1)} \mathcal{W}\Pi_5^{(1)} $ anti-commutes with the unitary operator $\Gamma_5^{(1)}\Gamma_5^{(2)}$. Since $\Gamma_5^{(1)}\Gamma_5^{(2)}$ has a structure like the Dirac matrix $\gamma_5$, $\Pi_5^{(1)} \mathcal{W}\Pi_5^{(1)} $ is chiral. Due to the Hermitcity of $\Pi_5^{(1)} \mathcal{W}\Pi_5^{(1)} $ the two off-diagonal blocks resulting from the chiral structure are related to each other. Therefore the Dirac operator is first of all doubly degenerate, namely it splits into twice the staggered Dirac-operator. Second, the off-diagonal blocks of $\Pi_5^{(1)} \mathcal{W}\Pi_5^{(1)} $ are either imaginary quaternion equivalent to real quaternion ($\beta=1$), complex ($\beta=2$) or purely imaginary equivalent to real ($\beta=4$). Thus they share the same universality classes as well as the same symmetry breaking pattern, see table~\ref{table2} as the staggered Dirac operator in four dimensions \cite{Stag,Osb10}. In Fig.~\ref{fig1} we compare some RMT predictions with lattice data via measuring the microscopic level density.

Due to the drastic change of symmetries the original Dyson index $\beta$ does not agree anymore with the level repulsion predicted by the RMT for the continuum theory. Hence one can ask for the corresponding joint probability density function (jpdf) of the eigenvalues from an RMT perspective. The jpdf plays a crucial role in analyzing the eigenvalue statistics and serves as a good starting point for deriving observables used for fitting the LECs. The jpdf of all symmetry classes (except for the GUE) discussed in this work can be written in a unified form, i.e.
\be
 p(\Lambda)\prod\limits_{1\leq j\leq n}d\lambda_j\propto|\Delta_{n}(\Lambda^2)|^{\beta_{\rm eff}}
\prod\limits_{1\leq j\leq n}\exp\left[-n\lambda_j^2\right]\lambda_j^\alpha d\lambda_j,
\label{joint-lat}
\ee
where we have drawn the random matrices from Gaussian ensembles. For the jpdf's of the GUE as well as of the GOE and the GSE one has to replace $\Delta_{n}(\Lambda^2)\to\Delta_{n}(\Lambda)$. Notice that the effective Dyson index does not always agree with the Dyson index identified in the continuum theory, cf. table~\ref{table2}. The exponent $\alpha$ is a generic repulsion from the origin crucially depending on the symmetry class. For example one can derive the microscopic level density from the jpdf~\eqref{joint-lat}. Examples of  comparisons of those level densities with the corresponding  lattice simulations are shown in Fig.~\ref{fig1}.

\section{Conclusions and Outlook}\label{conc}

We classified the two dimensional continuum theory of QCD-like theories as well as their naive discretization  along the ten-fold classification \cite{class} of random matrices. The quenched strong coupling lattice simulations at small volumes  show an agreement with RMT predictions comparable to the agreement found in three- and four-dimensional QCD with chiral RMT \cite{Stag}. The reason for this good agreement may lie in the non-compact supersymmetric space dual to the (partially)~quenched partition function generating the level density of the Dirac operator and, thus, the order parameter of the spontaneous breaking of chiral symmetry, the chiral condensate. Although the Mermin-Wagner-Coleman theorem states that a continuous symmetry cannot be spontaneously broken in two dimensions, it is quite controversial if this also applies to non-compact symmetries \cite{anal}.

 The ultimate goal of our investigation is a fundamental understanding of the lattice artefacts of staggered fermions which will be discussed in forthcoming publications. An RMT for  four-dimensional QCD with the fermions in the fundamental representation of $\SU(3)$ was already proposed in \cite{Osb10}. Since the model of \cite{Osb10} is quite cumbersome we hope to find some simplifications by studying the mechanism of changing the universality class due to lattice artefacts. Our approach serves as a good starting point for such an analysis and can be easily extended to higher dimensions.

{\bf Acknowledgments.}
We acknowledge support by the Alexander-von-Humboldt Foundation (MK) and by U.S. DOE Grant No. DE-FG-88ER40388 (JV and SZ). Moreover we thank 
Alexander Altland, Poul Damgaard, Nikita Nekrasov and Martin Zirnbauer for fruitful discussions.

\end{document}